\documentclass{article}

\usepackage{arxiv}

\usepackage[utf8]{inputenc} 
\usepackage[T1]{fontenc}    
\usepackage{hyperref}       
\usepackage{url}            
\usepackage{booktabs}       
\usepackage{amsfonts}       
\usepackage{nicefrac}       
\usepackage{microtype}      
\usepackage{lipsum}
\usepackage{graphicx}
\usepackage{algorithmic}
\usepackage[tight,footnotesize]{subfigure}
\usepackage{caption}
\usepackage{adjustbox}
\usepackage{makecell} 
\usepackage[linesnumbered,ruled]{algorithm2e}

\usepackage{pifont}
\newcommand{\cmark}{\ding{51}}%
\newcommand{\xmark}{\ding{55}}%

\graphicspath{ {./images/} }

\title{Model Context Contracts --- MCP-Enabled Framework to Integrate LLMs With Blockchain Smart Contracts}

\author{
Eranga Bandara \\
Old Dominion University \\
USA \\
\texttt{cmedawer@odu.edu} \\
\And
Sachin Shetty \\
Old Dominion University \\
USA \\
\texttt{sshetty@odu.edu} \\
\And
Ravi Mukkamala \\
Old Dominion University \\
USA \\
\texttt{mukka@odu.edu} \\
\And
Ross Gore \\
Old Dominion University \\
USA \\
\texttt{rgore@odu.edu} \\
\And
Peter Foytik \\
Old Dominion University \\
USA \\
\texttt{pfoytik@odu.edu} \\
\And
Safdar H. Bouk \\
Old Dominion University \\
USA \\
\texttt{sbouk@odu.edu} \\
\And
Abdul Rahman \\
Deloitte \& Touche LLP \\
USA \\
\texttt{abdulrahman@deloitte.com} \\
\And
Xueping Liang \\
Florida International University \\
USA \\
\texttt{xuliang@fiu.edu} \\
\And
Ng Wee Keong \\
Nanyang Technological University \\
Singapore \\
\texttt{AWKNG@ntu.edu.sg} \\
\And
Kasun De Zoysa \\
University of Colombo\\
Sri Lanka \\
\texttt{kasun@ucsc.cmb.ac.lk} \\
\And
Aruna Withanage -  \\
Effectz AI \\
\texttt{aruna@effectz.ai} \\
\And
Nilaan Loganathan -  \\
Effectz AI \\
\texttt{nilaan@effectz.ai} \\
}

\begin{document}
\maketitle
\begin{abstract}
In recent years, blockchain has experienced widespread adoption across various industries, becoming integral to numerous enterprise applications. Concurrently, the rise of generative AI and large language models (LLMs) has transformed human-computer interactions, offering advanced capabilities in understanding and generating human-like text. The introduction of the Model Context Protocol (MCP) has further enhanced AI integration by standardizing communication between AI systems and external data sources. Despite these advancements, there is still no standardized method for seamlessly integrating LLM applications and blockchain. To address this concern, we propose ``MCC: Model Context Contracts" a novel framework that enables LLMs to interact directly with blockchain smart contracts through MCP-like protocol. This integration allows AI agents to invoke blockchain smart contracts, facilitating more dynamic and context-aware interactions between users and blockchain networks. Essentially, it empowers users to interact with blockchain systems and perform transactions using queries in natural language. Within this proposed architecture, blockchain smart contracts can function as intelligent agents capable of recognizing user input in natural language and executing the corresponding transactions. To ensure that the LLM accurately interprets natural language inputs and maps them to the appropriate MCP functions, the LLM was fine-tuned using a custom dataset comprising user inputs paired with their corresponding MCP server functions. This fine-tuning process significantly improved the platform's performance and accuracy. To validate the effectiveness of MCC, we have developed an end-to-end prototype implemented on the Rahasak blockchain with the fine-tuned Llama-4 LLM. This implementation demonstrates the practical feasibility of MCC and paves the way for future advances in AI-blockchain integration. To the best of our knowledge, this research represents the first approach to using the concept of Model Context Protocol to integrate LLMs with blockchain.
\end{abstract}


\section{Introduction}

Blockchain has transformed various industries by providing decentralized, transparent, and secure platforms for transactions and data management~\cite{rahasak, tikiri}. A significant innovation within this domain is the advent of smart contracts—self-executing agreements with terms directly embedded in code~\cite{aplos}. These contracts automatically enforce and execute agreements when predefined conditions are met, eliminating the need for intermediaries and enhancing operational efficiency. Concurrently, the field of artificial intelligence (AI) has witnessed remarkable advancements, particularly with the emergence of LLMs. These models, exemplified by systems like OpenAI's GPT series~\cite{llm, gpt-llm}, have transformed human-computer interactions through their ability to comprehend and generate human-like text. The widespread adoption of LLMs across sectors such as healthcare, finance, and education underscores their versatility and impact. The introduction of the Model Context Protocols (MCP) has further enhanced AI integration by standardizing communication between AI systems and external data sources~\cite{mcp}. MCP provides a universal, open standard for connecting AI systems with data sources, replacing fragmented integrations with a single protocol.

Despite the individual successes of blockchain and LLM technologies, integrating them remains a complex challenge due to differences in architecture and communication protocols~\cite{cllm}. To address this concern, we propose ``MCC: Model Context Contracts" a novel framework that enables LLMs to interact directly with blockchain smart contracts through MCP. This framework enables LLMs to directly interact with blockchain networks, allowing users to execute smart contract functions through natural language queries. By translating user inputs into actionable commands, MCC enhances the accessibility and usability of blockchain systems. To ensure that the LLM accurately interprets natural language inputs and maps them to the appropriate MCP functions, we employed a LLM fine-tuning approach~\cite{llm-finetune, mistral-fine-tune}. The LLM was fine-tuned using a custom dataset comprising user inputs (e.g., transaction requests and queries in natural language format) paired with their corresponding MCP server functions~\cite{llamafactory-unsloth}. This fine-tuning process significantly improved the performance and accuracy of the platform, enabling the LLM to effectively determine which MCP server function to invoke based on user input.

To validate the efficacy of the MCC framework, we developed a comprehensive prototype built on the Rahasak blockchain~\cite{rahasak}, integrated with a fine-tuned Llama-4 LLM~\cite{llama-4, bassa-llama}. Rahasak's scalable, microservices-based architecture—combined with its functional programming and actor-based Aplos smart contract~\cite{aplos, saas, indy} framework—provides an optimal environment for implementing and evaluating the MCC approach. This infrastructure ensures robust, modular, and efficient interactions between AI models and blockchain smart contracts. This paper presents the design and implementation of the MCC framework, demonstrating its potential to seamlessly bridge LLMs with blockchain functionality. The main contributions of this work are outlined as follows.

\begin{enumerate}
    \item Proposed a novel protocol for LLMs to interact with blockchain smart contracts through the Model Context Protocol(MCP).
    \item Developed a novel mechanism enabling users to interact with blockchain and perform transactions using natural language queries, utilizing LLM agents and MCP.
    \item Fine-tuned Meta's Llama-4 LLM to efficiently identify and map potential MCP functions based on user natural language queries to perform transactions.
    \item Implemented a prototype of the platform using the Rahasak blockchain and the fine-tuned Llama-4 model.
\end{enumerate}

\begin{figure}[h]
\centering{}
\vspace{0.1in}
\includegraphics[width=4.5in]{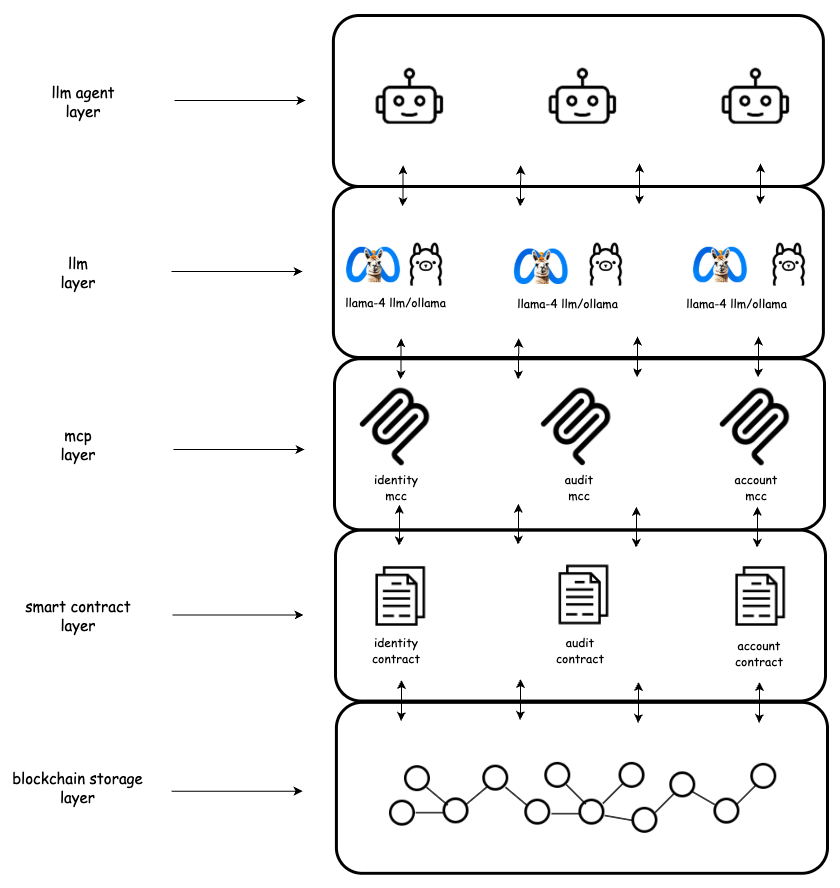}
\DeclareGraphicsExtensions.
\caption{MCC framework layered architecture.}
\label{indy528-architecture}
\end{figure}

\section{System Architecture}

Figure~\ref{indy528-architecture} describes the architecture of the platform. The proposed platform is composed of 5 layers: 1) Blockchain Storage Layer, 2) Smart Contract Layer, 3) MCP Layer, 4) LLM Layer, 5) LLM Agent Layer. Below is a brief description of each layer.

\subsection{Blockchain Storage Layer}

The Blockchain Storage Layer serves as the foundational component responsible for maintaining all blockchain-related data, including transaction records and digital assets. This layer ensures the integrity, immutability, and availability of data across the decentralized network~\cite{rahasak, hyperledger}. In the context of the Rahasak blockchain platform, the underlying storage mechanism utilizes Apache Cassandra, a highly scalable and distributed NoSQL database. Cassandra's architecture is particularly well-suited for high-throughput environments, supporting up to one million writes per second, which is ideal for applications requiring rapid transaction processing and scalability. Within this architecture, smart contracts interact directly with the asset storage system. Based on transaction inputs, smart contracts perform read and write operations on the asset storage, enabling dynamic and automated management of blockchain assets~\cite{aplos}. This integration ensures that all asset-related transactions are executed efficiently and recorded securely within the blockchain's storage infrastructure.

\subsection{Smart Contract Layer}

The Smart Contract Layer serves as the operational core of the blockchain platform, enabling the automation and enforcement of agreements without intermediaries. Smart contracts are self-executing programs stored on the blockchain that automatically execute when predefined conditions are met, ensuring transparency, security, and efficiency in transactions~\cite{saas}. In this layer, smart contracts interact directly with the Blockchain Storage Layer to perform read and write operations. Based on transaction inputs, they access and modify data related to digital assets and transaction records. For instance, in the Rahasak blockchain platform, smart contracts utilize Apache Cassandra as the underlying asset storage system. This integration allows smart contracts to efficiently manage and update asset states, ensuring that all transactions are accurately recorded and maintained within the decentralized ledger. By facilitating direct interaction with the storage layer, the Smart Contract Layer ensures that contractual agreements are executed precisely as coded, enhancing trust and reducing the potential for disputes among parties.

\subsection{MCP Layer}

The MCP Layer serves as a standardized interface that facilitates seamless communication between LLMs and external tools, data sources, and services. By implementing MCP, AI applications can dynamically access and interact with various resources, enhancing their functionality and contextual understanding~\cite{mcp}. In the MCC framework, the MCP Layer facilitates seamless interaction between LLMs and blockchain smart contracts. The MCP Layer enables LLMs to invoke specific functions exposed by MCP-compliant servers, which act as intermediaries connecting the LLM with blockchain functionalities, including smart contract operations. Each function in the MCP server corresponds directly to a smart contract function on the blockchain, establishing a one-to-one mapping. These functions are identified as Model Context Contracts (MCCs), serving as standardized interfaces that allow LLMs to execute blockchain operations through natural language queries. This setup allows users to interact with the blockchain using natural language inputs, which are processed by the LLM and translated into corresponding MCP function calls~\cite{mcp-safety}. The MCP Layer's client-server architecture ensures flexibility and scalability, allowing the integration of multiple external services and data sources without the need for custom integrations. This standardization simplifies the development process and enhances the AI application's ability to perform complex tasks by leveraging a diverse set of tools and data.

\subsection{LLM Layer}

The LLM Layer serves as the cognitive engine of the Model Context Contracts (MCC) framework, enabling the interpretation and generation of human-like language. This layer leverages advanced deep learning models, such as transformer-based architectures, to process and understand natural language inputs from users~\cite{llm}. By analyzing these inputs, the LLM generates coherent and contextually relevant responses, facilitating seamless communication between users and the system.

In the MCC framework, the LLM Layer plays a pivotal role in bridging user interactions with blockchain functionalities. When a user submits a natural language query or command, the LLM interprets the input to discern the user's intent and the specific actions required. This interpretation involves parsing the input, understanding contextual nuances, and determining the appropriate blockchain operations to perform.

To enhance the accuracy and relevance of these interpretations, the LLM is fine-tuned using a custom dataset comprising pairs of natural language inputs and their corresponding MCP server functions~\cite{mistral-fine-tune, devsec-gpt}. Through this fine-tuning process, the LLM learns to map user inputs to specific MCP functions effectively, thereby improving the precision of function invocation and the overall responsiveness of the system.

By integrating the LLM Layer into the platform, users can interact with the blockchain system using intuitive natural language, eliminating the need for complex command syntax or specialized knowledge. This advancement significantly enhances user accessibility and engagement, making blockchain operations more approachable and user-friendly.

\begin{figure}[h]
\centering{}
\vspace{0.1in}
\includegraphics[width=5.2in]{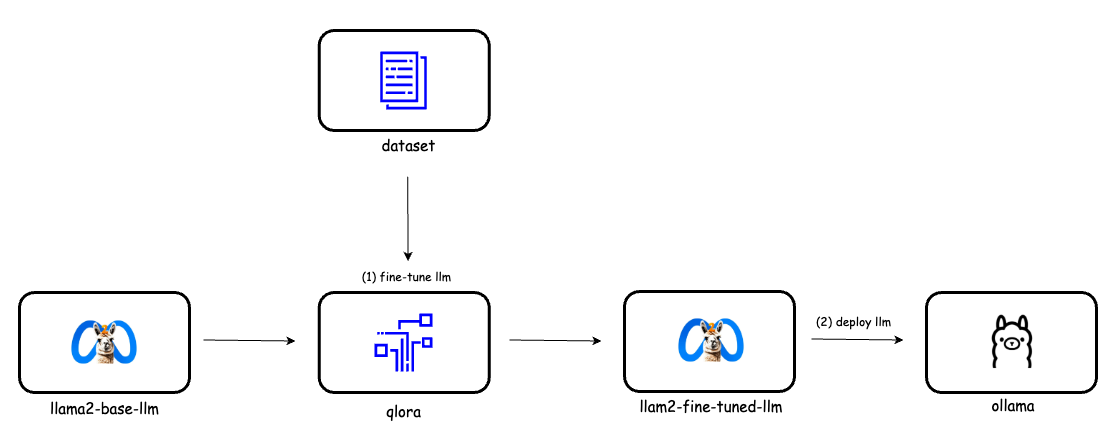}
\DeclareGraphicsExtensions.
\caption{Fine-tune LLMs with Qlora and deploy with Ollama.}
\label{llm-fine-tune}
\end{figure}

\begin{figure}[t]
\centering{}
\includegraphics[width=5.2in]{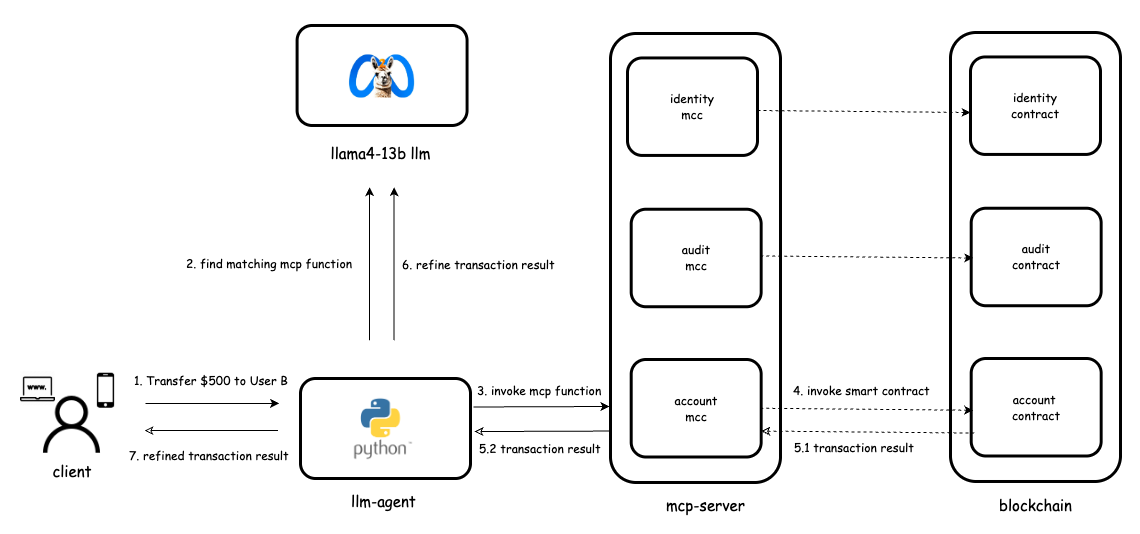}
\DeclareGraphicsExtensions.
\caption{End-to-end interaction between the client, LLM agent, fine-tuned LLM, MCC functions, and the underlying blockchain.}
\label{llama2-flow}
\end{figure}

\subsection{LLM Agent Layer}

The LLM Agent Layer serves as the orchestration hub within the MCC framework, enabling seamless and secure interactions in decentralized environments. It begins by capturing user-submitted transaction requests or queries expressed in natural language and forwarding them to the LLM Layer for interpretation~\cite{agentic-ai}. Upon receiving the processed input, the LLM Agent Layer determines the appropriate MCP function that aligns with the user's intent, ensuring the correct blockchain operation is identified for execution.

Subsequently, it invokes the identified MCP function, which interacts with the corresponding smart contract function on the blockchain, thereby executing the user's request accurately within the network. To uphold transaction integrity and security, the LLM Agent Layer manages essential cryptographic operations, including the digital signing of transactions, by securely handling private keys to authenticate and authorize all transactions~\cite{aplos, pop}.

In decentralized settings, this layer is designed to operate in close proximity to the client, such as within a user's mobile wallet application, enhancing responsiveness and reducing latency for efficient and immediate blockchain interactions. By integrating these functionalities, the LLM Agent Layer empowers users to engage with blockchain systems using intuitive natural language, effectively bridging the gap between complex blockchain operations and user-friendly interfaces.

\section{Platform Functionality}

The platform is built around five core functionalities that work in concert to enable seamless interaction between users and blockchain systems through natural language. These include: (1) fine-tuning of LLMs, (2) accepting user inputs for blockchain transactions in the form of natural language queries, (3) identifying the appropriate MCP functions via the LLM, (4) invoking the corresponding smart contract functions through the MCP server, and (5) handling and refining responses from the blockchain. This section details each of these components through a representative use case, where an account smart contract on the blockchain supports two operations: `transfer\_funds` and `get\_account\_balance`. Figure~\ref{llama2-flow} illustrates the end-to-end interaction between the different layers of the system as they work together to fulfill this use case.

\subsection{LLM Fine-Tuning}

Fine-tuning the LLM is a crucial component of the platform, enhancing its ability to interpret user inputs and accurately map them to corresponding MCP functions. This process involves adapting a pre-trained LLM to the specific task of function calling by training it on a custom dataset~\cite{llm-finetune, mistral-fine-tune}. The dataset comprises pairs of natural language user queries—such as transaction requests—and their respective MCP function calls. By exposing the LLM to these examples, it learns to discern patterns and relationships between user intents and the appropriate blockchain operations. This targeted training significantly improves the model's performance in translating diverse natural language inputs into precise function calls, thereby reducing errors and enhancing the overall efficiency of the system. To optimize resource efficiency and enable edge deployment, the LLM fine-tuning was performed using the Unsloth library~\cite{llamafactory-unsloth} with Low-Rank Adapters (LoRA) and 4-bit quantization (QLoRA)~\cite{qlora, slice-gpt}, ensuring high performance with minimal memory overhead. These quantized models are then deployed via Ollama, a lightweight framework for high-performance vision-language inference, Figure~\ref{llm-fine-tune}. The outcome is a robust LLM capable of facilitating seamless and intuitive user interactions with the blockchain through natural language.

\subsection{User Input of Blockchain Transactions via Natural Language Queries}

In the MCC framework, the integration of LLMs has transformed user interactions with blockchain systems by enabling natural language processing capabilities. The LLM Agent Layer plays a crucial role in facilitating user-initiated blockchain transactions through conversational language~\cite{mcp-safety}. When a user inputs a transaction request, such as ``What is my current balance?" or ``Transfer \$500 to user\_2," the LLM Agent captures this input and forwards it to the LLM for interpretation. The LLM processes the natural language input to comprehend the user's intent and determines the appropriate MCP function that corresponds to the requested blockchain operation. Upon identifying the suitable MCP function, the LLM Agent invokes it, which subsequently interacts with the relevant smart contract function on the blockchain to execute the transaction. This seamless process allows users to perform complex blockchain operations without needing to understand the underlying technicalities. Moreover, the LLM Agent Layer manages essential cryptographic operations, such as digitally signing transactions, ensuring the security and authenticity of each operation~\cite{llm-agents}. Strategically residing close to the client, for instance, within a user's mobile wallet application, the LLM Agent Layer enhances responsiveness and reduces latency, providing efficient and immediate blockchain interactions in a decentralized environment. This architecture not only simplifies user engagement with blockchain systems but also upholds the integrity and security of transactions.

\subsection{LLM Identification of MCP Functions}

In the MCC framework, the fine-tuned LLM serves as a critical bridge between user inputs and blockchain functionalities. When a user submits a transaction request or query in natural language, the LLM Agent Layer captures this input and forwards it to the LLM for processing. Leveraging its fine-tuned capabilities, the LLM interprets the user's intent and accurately identifies the corresponding MCP function required to execute the desired blockchain operation~\cite{mistral-fine-tune}. For instance, if user\_1 inquires about their balance, the LLM discerns that the 'get\_account\_balance' MCP function should be invoked. Similarly, if user\_1 requests to transfer funds to user\_2, the LLM identifies the 'transfer\_funds' MCP function as appropriate. This precise mapping ensures that the appropriate MCP function is invoked, which subsequently interacts with the relevant smart contract on the blockchain to complete the transaction. The fine-tuning process enhances the LLM's proficiency in understanding domain-specific terminology and user intents, thereby improving the reliability and efficiency of function calling within the system~\cite{qlora}. By effectively translating natural language inputs into actionable blockchain commands, the fine-tuned LLM significantly enhances user interaction with the blockchain platform, making it more intuitive and accessible.

\subsection{MCP Server Invocation of Smart Contract Functions}

In the MCC framework, the MCP server plays a pivotal role in facilitating interactions between the LLM Agent and blockchain smart contracts. Once the LLM Agent determines the appropriate MCP function corresponding to a user's natural language query, it invokes this function on the MCP server, supplying the necessary arguments and transaction parameters. The MCP server then processes this invocation by mapping the specified function to the relevant smart contract method on the blockchain~\cite{rahasak, aplos}. For example, if the 'transfer\_funds' MCP function is invoked, the MCP server interacts with the smart contract's 'transfer' function, executing the fund transfer from user\_1 to user\_2. Subsequently, the server executes the smart contract function, effectively carrying out the user's intended transaction within the blockchain network. This seamless orchestration ensures that complex blockchain operations can be performed through intuitive natural language inputs, thereby enhancing user accessibility and interaction with blockchain systems~\cite{agentic-ai}. Moreover, the MCP server manages essential cryptographic operations, such as digitally signing transactions, to uphold the security and integrity of each operation. By facilitating this streamlined process, the MCP server enables efficient and secure execution of blockchain transactions initiated through natural language commands. 

\subsection{Handle Response From the Blockchain}

In the MCC framework, effectively managing and conveying blockchain responses is essential for providing users with clear and actionable feedback regarding their transactions. After a smart contract executes a transaction—such as processing a fund transfer from user\_1 to user\_2—the MCP server retrieves the outcome and transmits this transaction response back to the LLM Agent~\cite{llm-agents}. The LLM Agent then forwards this response as context to the LLM, which processes and refines the information into an understandable and user-friendly explanation. Finally, the LLM delivers this refined response to the user, providing a clear summary of the transaction's result. This process enhances user experience by translating complex blockchain responses into easily digestible information, thereby bridging the gap between intricate blockchain operations and user-friendly interactions~\cite{agentic-ai}.

\begin{figure}[h]
\centering{}
\vspace{0.1in}
\includegraphics[width=5.5in]{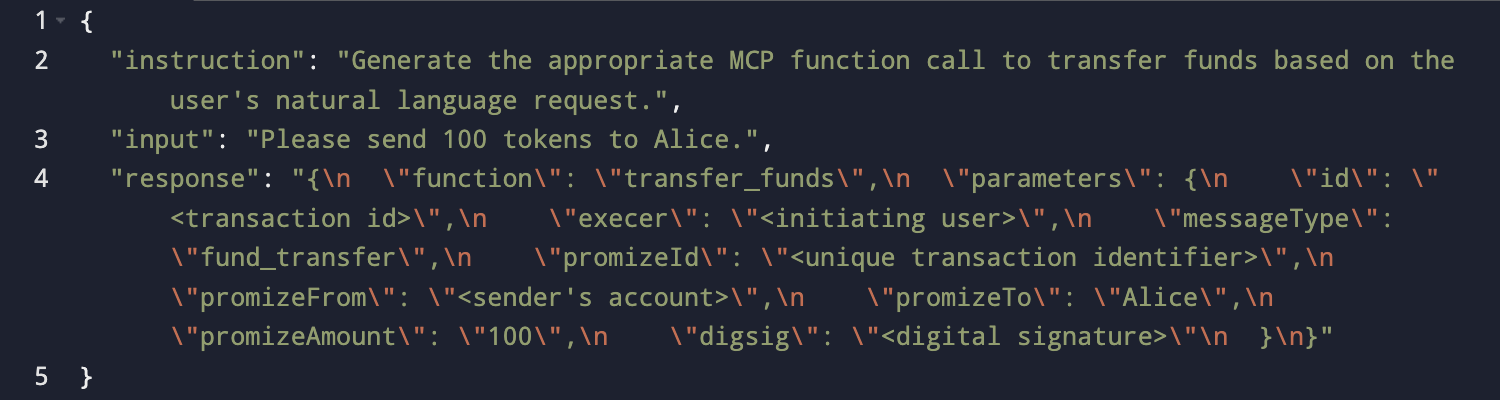}
\DeclareGraphicsExtensions.
\caption{The required data format of the unsloth library to fine-tune the Llama-4 LLM.}
\label{unsloth-format}
\end{figure}

\section{Implementation and Evaluation}

A functional prototype of the MCC framework has been implemented using the Rahasak blockchain and its Aplos smart contracts. Rahasak introduces a novel Validate-Execute-Group blockchain architecture, wherein transactions are executed a single time on a peer, enhancing efficiency and scalability~\cite{rahasak, casper}. The LLM Layer is constructed with a fine-tuned Llama-4 model~\cite{llama-4}. The fine-tuning process utilized synthetically generated datasets comprising user natural language queries and their respective MCP functions and parameters. Each dataset entry was paired with a structured JSON representation to facilitate effective learning. Fine-tuning was conducted using the Unsloth library on Google Colab, leveraging NVIDIA A100 and Tesla TPU resources for accelerated training cycles~\cite{mistral-fine-tune}. As Unsloth requires datasets in a conversational instruction-tuning format, annotations were transformed into structured conversation prompts~\cite{prompt-engineering}. The structure of the dataset used to fine-tuned the LLM presented in the Figure~\ref{unsloth-format}. Each entry contained a content field (the user's natural language query), type (the nature of the query—e.g., balance inquiry, fund transfer), and an instruction guiding the model on how to map the query to the appropriate MCP function and parameters. Post fine-tuning, models were optimized using Quantized Low-Rank Adaptation (QLoRA), allowing quantized deployment on edge or resource-constrained hardware~\cite{qlora}. These models were then hosted on the Ollama framework, providing an efficient runtime environment for inference~\cite{ollama}. This implementation demonstrates the practical feasibility of integrating fine-tuned LLMs with blockchain systems, enabling users to interact with the blockchain using natural language queries effectively. The platform’s performance is evaluated across two key areas: Evaluation of LLM and Evaluation of the blockchain.

\begin{figure}[h]
\centering{}
\vspace{0.1in}
\includegraphics[width=5.0in]{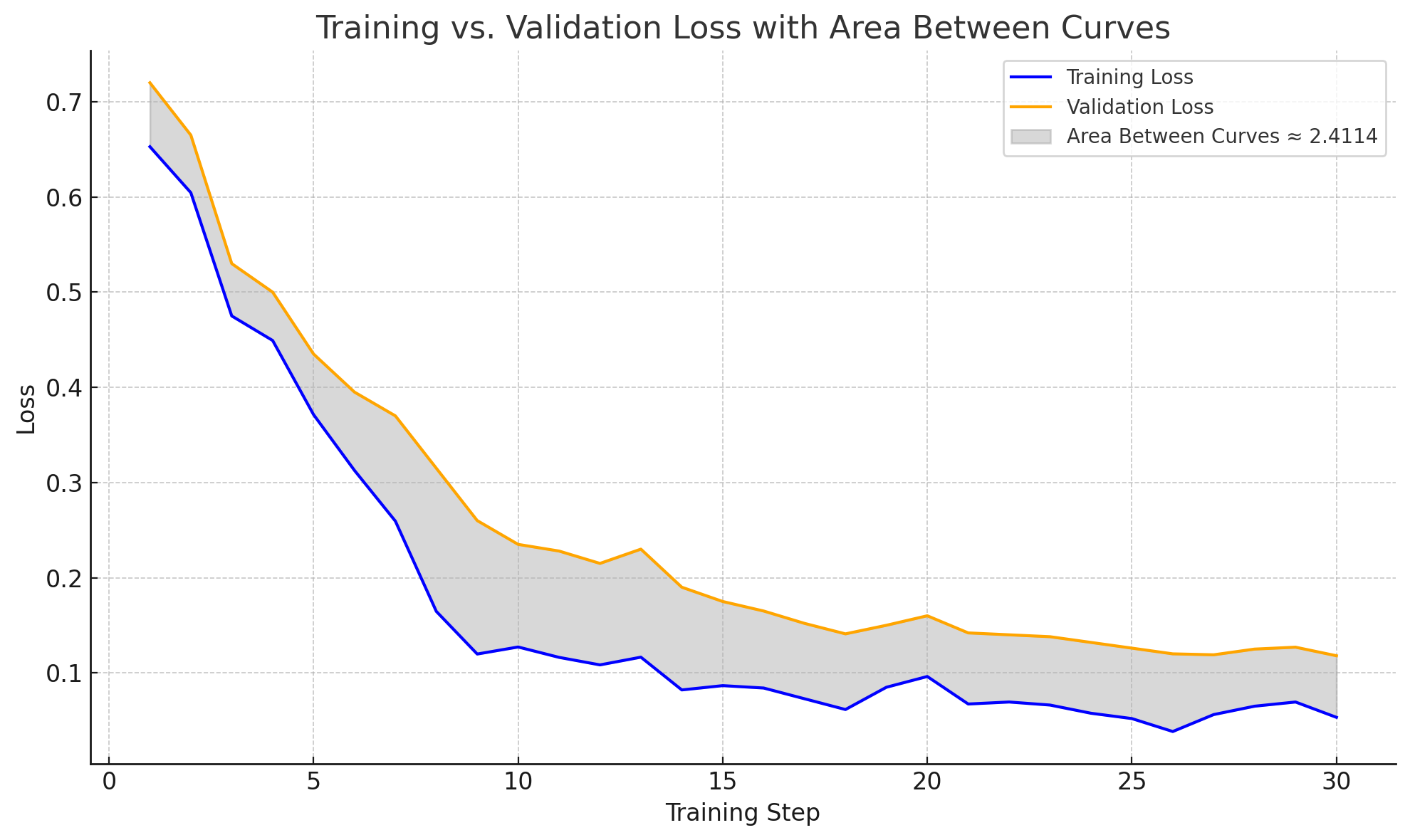}
\DeclareGraphicsExtensions.
\caption{Training loss and validation loss during fine-tuning of the Llama-4 LLM.}
\label{unsloth-tranning-validation-loss}
\end{figure}

\begin{figure}[h]
\centering{}
\vspace{0.1in}
\includegraphics[width=5.0in]{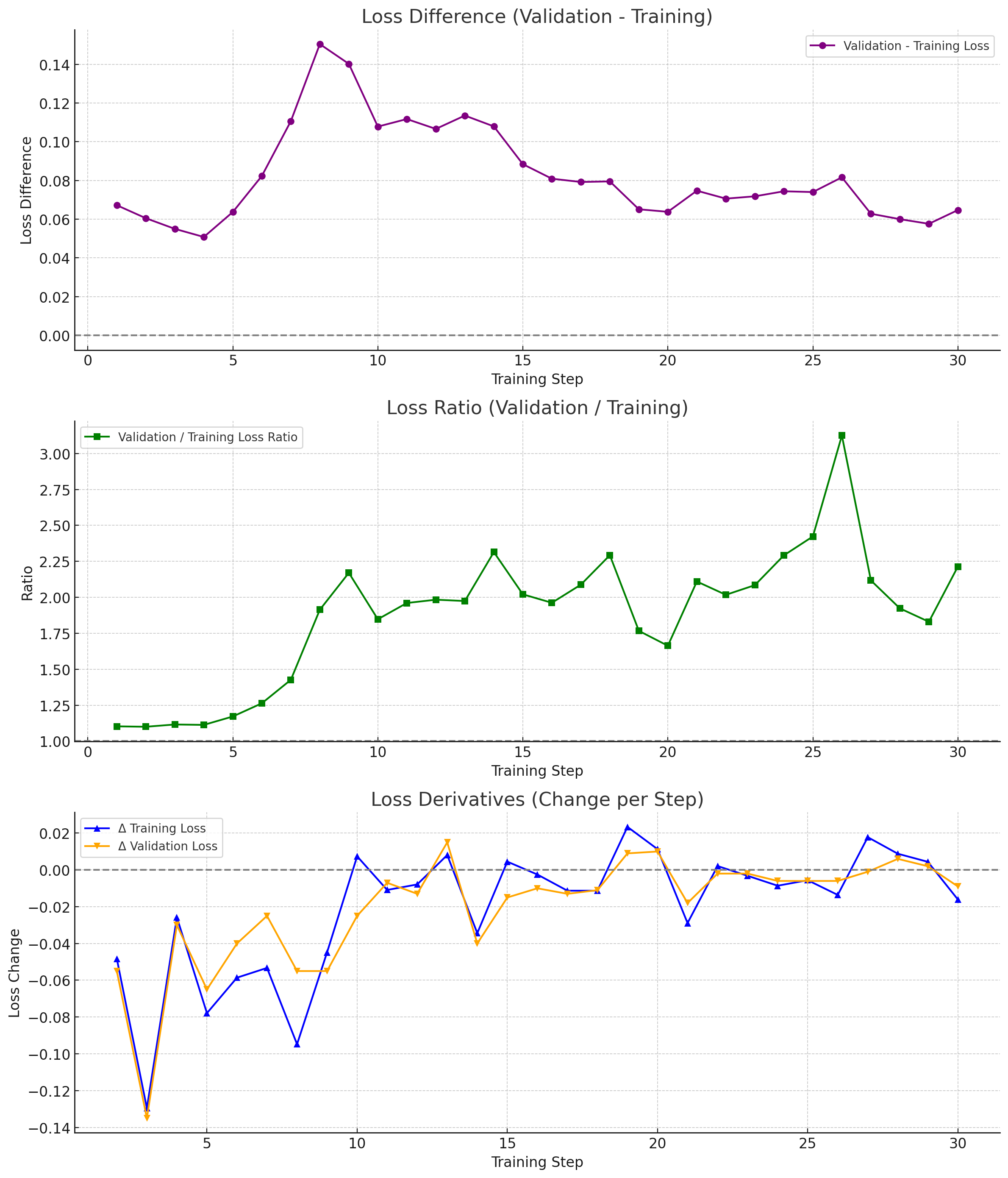}
\DeclareGraphicsExtensions.
\caption{Ratio of training to validation loss during the fine-tuning of the Llama-4.2-11B-Vision-Instruct vision-language model.}
\label{unsloth-loss-ratio}
\end{figure}

\subsection{Evaluation of LLM}

In evaluating the fine-tuned Llama-4 LLM within the MCC framework, we focused on its ability to accurately interpret user inputs and map them to appropriate MCP functions with corresponding parameters. During the fine-tuning process, we monitored training and validation loss metrics to assess the model's learning progression~\cite{mistral-fine-tune}. These metrics, visualized in Figure~\ref{unsloth-tranning-validation-loss}, demonstrate the model's progressive learning across training steps. Additionally, Figure~\ref{unsloth-loss-ratio} captures the rate of change in loss, offering insights into the convergence behavior and training stability of the model.

\begin{figure}[h]
\centering{}
\vspace{0.1in}
\includegraphics[width=5.2in]{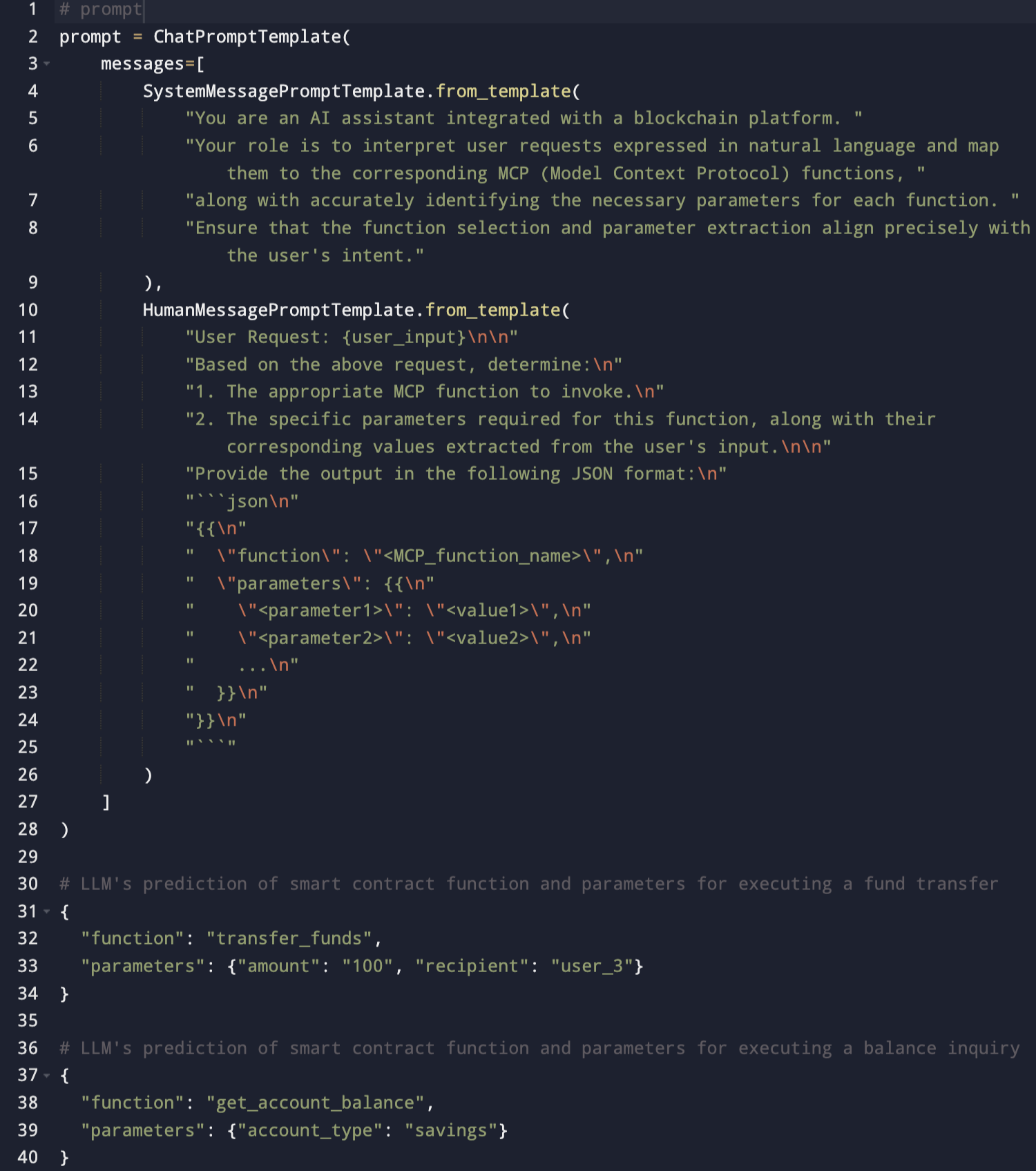}
\DeclareGraphicsExtensions.
\caption{Prompt used to guide the LLM in predicting the appropriate smart contract function and corresponding parameters, along with the LLM’s generated response.}
\label{prompt}
\end{figure}

Post fine-tuning, we evaluated the LLM's predictive performance using a curated validation set comprising unseen user queries related to blockchain operations. For instance, when presented with a user request such as ``Transfer 100 tokens to user\_3," the model accurately identified the corresponding MCP function transfer\_funds and extracted the parameters amount: 100 and recipient: user\_3. Similarly, for a query like ``Check my savings account balance," the model correctly mapped it to the get\_account\_balance function with the parameter account\_type: savings. These examples, detailed in Figure~\ref{prompt}, showcase the model's capability to deliver precise and actionable insights. The results confirm that the fine-tuned LLM effectively translates natural language queries into structured MCP function calls with appropriate parameters, enhancing the MCC framework's ability to facilitate intuitive and accurate user interactions with blockchain systems.



\subsection{Evaluation of Blockchain}

\begin{figure}[h]
\centering{}
\includegraphics[width=5.2in]{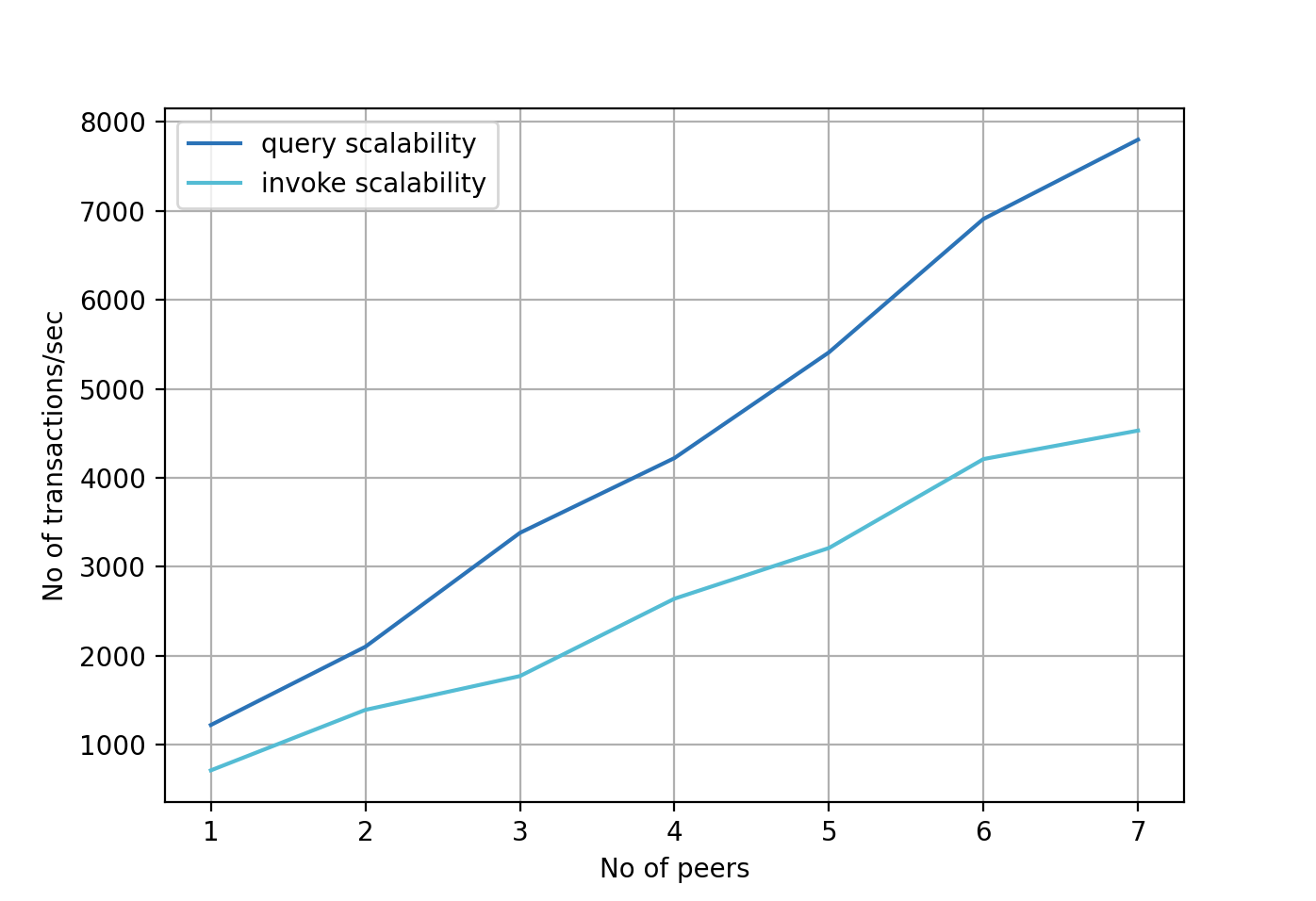}
\DeclareGraphicsExtensions.
\caption{Transaction scalability}
\label{tr-scale}
\end{figure}

\begin{figure}[h]
\centering{}
\includegraphics[width=5.2in]{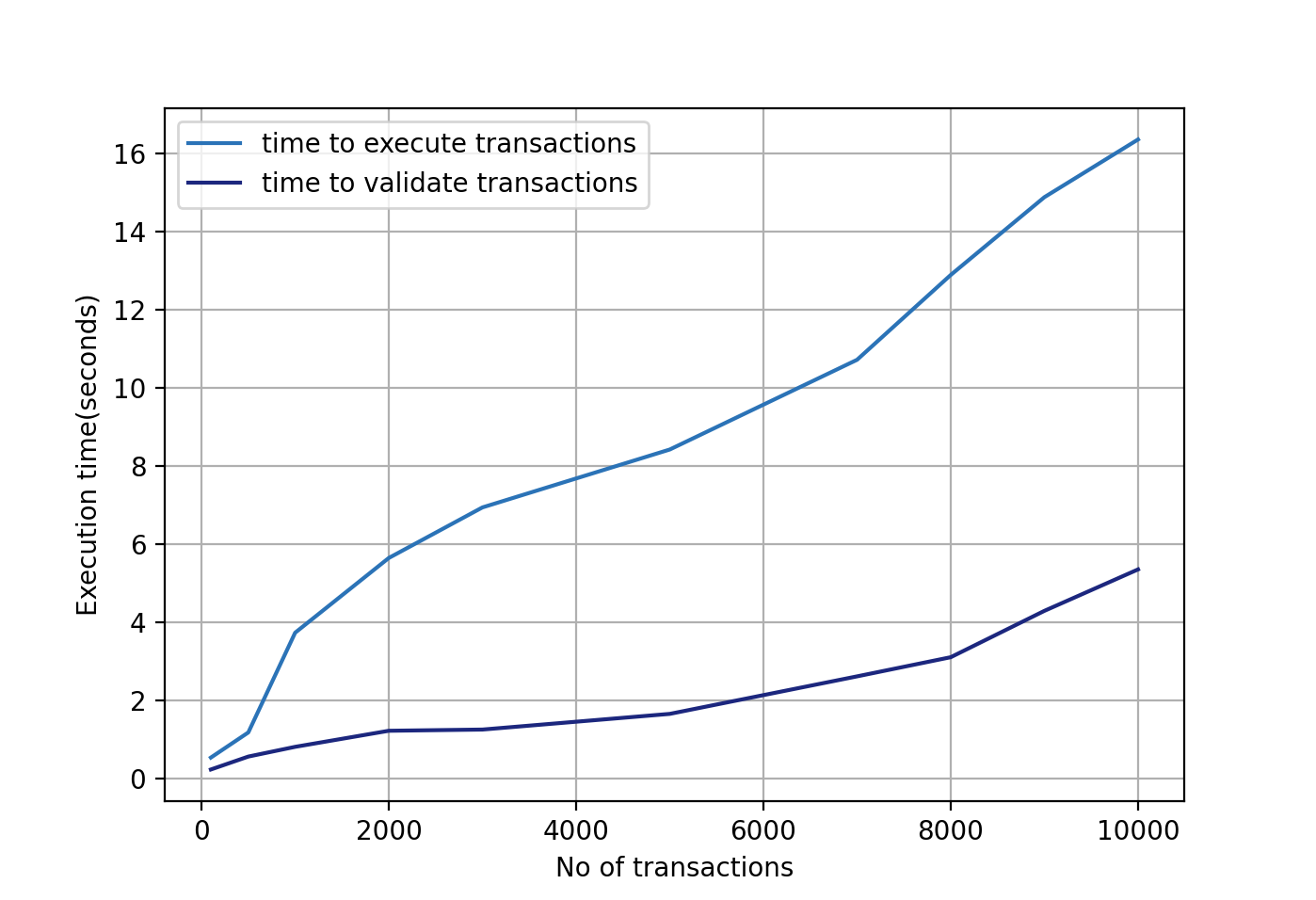}
\DeclareGraphicsExtensions.
\caption{Time to execute/validate transactions.}
\label{tr-exec}
\end{figure}

\begin{figure}[h]
\centering{}
\includegraphics[width=5.2in]{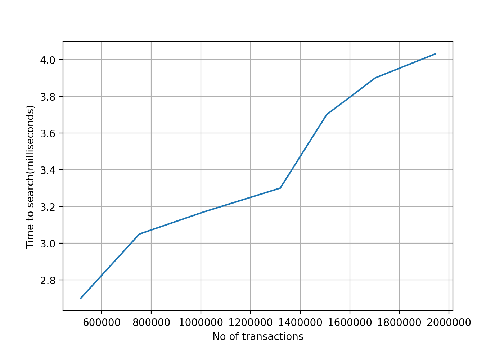}
\DeclareGraphicsExtensions.
\caption{Transaction search throughput of blockchain.}
\label{tr-search}
\end{figure}

To evaluate the blockchain performance within the context of the MCC framework, we conducted experiments across varying numbers of blockchain peers. The focus was on assessing transaction scalability and the validation/execution efficiency of the underlying blockchain, specifically in response to requests initiated through the MCC server. We measured the throughput for both invoke transactions (e.g., `transfer\_funds`) and query transactions (e.g., `get\_account\_balance`), and observed linear scalability—throughput increased proportionally with the number of peers, as shown in Figure~\ref{tr-scale}. Transaction validation and execution times were also evaluated, including key operations such as digital signature verification, double-spend detection, smart contract execution, and ledger updates~\cite{rahasak}. The results, presented in Figure~\ref{tr-exec}, confirm consistent and efficient execution across an increasing network size. Additionally, the platform's search performance was evaluated using Rahasak’s Lucene-based index API for retrieving account-related data such as balances. As illustrated in Figure~\ref{tr-search}, the system demonstrated low-latency access, making it well-suited for real-time interactions within MCC-driven workflows.

\begin{table*}[!htb]\centering
\vspace{0.1in}
\caption {Comparison of MCC framework with Related Works}
\begin{adjustbox}{width=1\textwidth}
\label{t_mcc_comparison}
\begin{tabular}{lccccc}
\toprule
\thead{Platform} & \thead{Use case} & \thead{Running Blockchain} & \thead{Running LLM} & \thead{Fine-tuning\\Support} & \thead{MCP\\Support} \\
\midrule
MCC & LLM integration with Blockchain & Rahasak  & Llama-4 & \cmark & \cmark \\
C-LLM \cite{cllm} & LLM integration with Blockchain & Not specified & Various LLMs & \cmark & \xmark \\
NLP-Blockchain \cite{nlp-blockchain} & Smart Contract Generation & Not specified & Not specified & \cmark & \xmark \\
LLM-IPFS \cite{nlp-ipfs} & Academic Document Integrity & Not specified & BERT & \xmark & \xmark \\
Sligpt \cite{sligpt} & Data Dependency Analysis & Ethereum & GPT-4o & \xmark & \xmark \\
SCLA \cite{scla} & Smart Contract Summarization & Not specified & Gemini-1.5-Pro & \xmark & \xmark \\
LLM-Net \cite{llm-net} & Decentralized LLM Services & Not specified & Various LLMs & \cmark & \xmark \\
\bottomrule
\end{tabular}
\end{adjustbox}
\end{table*}

\section{Related Work}

Several researchers have explored the integration of blockchain technology with LLMs, aiming to enhance the functionality and intelligence of smart contracts. This section outlines key initiatives in this domain, highlighting their methodologies and architectures. Table~\ref{t_mcc_comparison} provides a comparative analysis of these approaches in relation to the proposed MCC framework.

Connecting Large Language Models with Blockchain Advancing the Evolution of Smart Contracts from Automation to Intelligence~\cite{chatgpt-llm} - This paper introduces the C-LLM framework, aiming to bridge the interoperability gap between LLMs and blockchain systems. By integrating semantic relatedness with truth discovery methods through the SenteTruth approach, it enhances the accuracy and trustworthiness of data generated by LLMs for smart contracts.

Smart Contract Generation through NLP and Blockchain for Legal Documents~\cite{nlp-blockchain} - This study focuses on automating the generation of smart contracts from legal documents using Natural Language Processing (NLP) techniques. By analyzing legislative texts, it facilitates the creation of intelligent code, streamlining the process of translating legal agreements into executable smart contracts on the blockchain.

Integrating LLM with Blockchain and IPFS to Enhance Academic Diploma Integrity~\cite{nlp-ipfs} - This research presents a solution to combat academic document fraud by combining LLMs with blockchain and the InterPlanetary File System (IPFS). The approach involves pre-validating academic documents using LLMs before their certification, ensuring the integrity and authenticity of academic diplomas.

Sligpt~\cite{sligpt}: A Large Language Model-Based Approach for Data Dependency Analysis on Solidity Smart Contracts - Sligpt integrates a LLM with the static analysis tool Slither to perform data dependency analyses on Solidity smart contracts. This methodology enhances the testing and security of smart contracts by leveraging the code comprehension capabilities of LLMs alongside traditional analysis tools.

SCLA~\cite{scla}: Automated Smart Contract Summarization via LLMs and Control Flow Prompt - The SCLA framework leverages LLMs and semantic augmentation to improve smart contract code summarization. By constructing Abstract Syntax Trees (ASTs) to extract latent semantics, it forms semantically enriched prompts, enhancing the LLM's understanding and summarization of smart contract code.

LLM-Net~\cite{llm-net}: Democratizing LLMs-as-a-Service through Blockchain-based Expert Networks - LLM-Net proposes a decentralized network of specialized LLM providers, utilizing blockchain to democratize LLMs-as-a-Service. It incorporates fine-tuned expert models for various domains, ensuring sustained knowledge growth and maintaining service quality through collaborative prompting mechanisms and transparent transaction validation.

\section{Conclusions and Future Work}

In this paper, we introduced the Model Context Contracts(MCC) framework, a novel approach that seamlessly integrates LLMs with blockchain smart contracts. By leveraging the MCP, MCC framework enables users to interact with blockchain systems through natural language queries, thereby enhancing accessibility and usability. The fine-tuning of Meta's Llama-4 LLM was instrumental in accurately mapping user inputs to corresponding MCP functions, ensuring precise execution of blockchain operations. Our implementation on the Rahasak blockchain demonstrated the framework's scalability and efficiency, underscoring its potential to bridge the gap between complex blockchain functionalities and user-friendly interfaces. This work contributes to the evolving landscape of AI and blockchain integration, paving the way for more intelligent and responsive decentralized applications. To the best of our knowledge, this research represents the first approach of using the Model Context Protocol concept to integrate LLMs with blockchain. For future work, we plan to expand the platform's capabilities by incorporating more open-source LLMs and adopting parameter-efficient fine-tuning approaches.




\bibliographystyle{unsrt}
\bibliography{reference}

\end{document}